%% file: main.tex
\documentclass[10pt,conference]{IEEEtran}
\usepackage{mathptmx} 

\usepackage{fancyhdr}
\usepackage[normalem]{ulem}
\usepackage[hyphens]{url}
\usepackage[sort,nocompress]{cite}
\usepackage[final]{microtype}
\usepackage[keeplastbox]{flushend}
\usepackage[bookmarks=true,breaklinks=true,letterpaper=true,colorlinks,citecolor=blue,linkcolor=blue,urlcolor=blue]{hyperref}
\usepackage{tikz}
\usepackage{xcolor}
\usepackage[normalem]{ulem}
\usepackage{amsmath}
\usepackage{multirow}
\usepackage{diagbox}
\usepackage{makecell}
\usepackage[capitalise]{cleveref}

\newcommand{\circuitname}{ESM}
\newcommand{\frameworkname}{QECC-Synth}
\newcommand{\qmrname}{SATmap}
\newcommand{\sabrename}{Sabre}
\newcommand{\surfname}{Surf-Stitch}

\newcommand{\phaseonename}{Initialization}
\newcommand{\phasetwoname}{Encoding}
\newcommand{\phasethreename}{Control}
\newcommand{\phasefourname}{Decoding}
\newcommand{\phasefivename}{Measurement}

\newcommand{\timeout}{Time-Limit}
\newcommand{\noexist}{Not-Exist}

\newcommand{\ballnumber}[1]{\tikz[baseline=(myanchor.base)] \node[circle,fill=.,inner sep=1pt] (myanchor) {\color{-.}\bfseries\footnotesize #1};
}

\usepackage{authblk}
\usepackage{titlesec}
\titleformat{\title}{\normalfont\LARGE\bfseries}{}{}{}
\title{\LARGE \frameworkname: A Layout Synthesizer for Quantum Error Correction Codes on Sparse Hardware Architectures}

\author[1]{Keyi Yin\thanks{keyin@ucsd.edu}}
\author[1]{Hezi Zhang\thanks{hez019@ucsd.edu}}
\author[1]{Xiang Fang\thanks{x8fang@ucsd.edu }}
\author[2]{Yunong Shi\thanks{shiyunon@amazon.com}}
\author[3]{Travis Humble\thanks{humblets@ornl.gov}}
\author[4]{Ang Li\thanks{ang.li@pnnl.gov}}
\author[1]{Yufei Ding\thanks{yufeiding@ucsd.edu}}

\affil[1]{University of California, San Diego, CA, USA}
\affil[2]{Amazon Braket, New York, NY, USA} 
\affil[3]{Oak Ridge National Laboratory, Oak Ridge, TN, USA}
\affil[4]{Pacific Northwest National Laboratory, Richland, WA, USA}

\begin{document}

\maketitle
\pagestyle{plain}

\input{01_abstract}
\input{02_introduction}

\input{03_background}

\input{04_formulation}

\input{05_tech}
\input{06_evaluation}
\input{07_related_work}

\input{08_conclusion}
\input{ACK}

\bibliographystyle{unsrt}
\bibliography{references}

\end{document}

%% file: 01_abstract.tex
\begin{abstract}
Quantum Error Correction (QEC) codes are essential for achieving fault-tolerant quantum computing (FTQC). However, their implementation faces significant challenges due to disparity between required dense qubit connectivity and sparse hardware architectures. Current approaches often either underutilize QEC circuit features or focus on manual designs tailored to specific codes and architectures, limiting their capability and generality. In response, we introduce \frameworkname, an automated compiler for QEC code implementation that addresses these challenges. We leverage the ancilla bridge technique tailored to the requirements of QEC circuits and introduces a systematic classification of its design space flexibilities. We then formalize this problem using the MaxSAT framework to optimize these flexibilities. Evaluation shows that our method significantly outperforms existing methods while demonstrating broader applicability across diverse QEC codes and hardware architectures. 
\end{abstract}

%% file: 02_introduction.tex
\section{Introduction}
\label{sec:introduction}

Current quantum devices \cite{Preskill2018QuantumCI, chow2021ibm, Holmes2020NISQBQ} suffer from errors, limiting their effectiveness in various applications~\cite{google2021exponential, Sycamore}. Quantum error correction (QEC) addresses this issue by encoding redundant physical qubits into logical qubits \cite{QCQI}. These logical qubits exhibit significantly lower error rates, ensuring the fidelity of quantum computations \cite{knill1998resilient, aharonov1997fault}. Recent advancements in small-scale QEC implementations \cite{marques2022logical, google2023suppressing, sivak2023real} demonstrate progress toward large-scale FTQC.

%

\begin{figure*}[!htbp]
    \centering
    \includegraphics[width=0.92\textwidth]{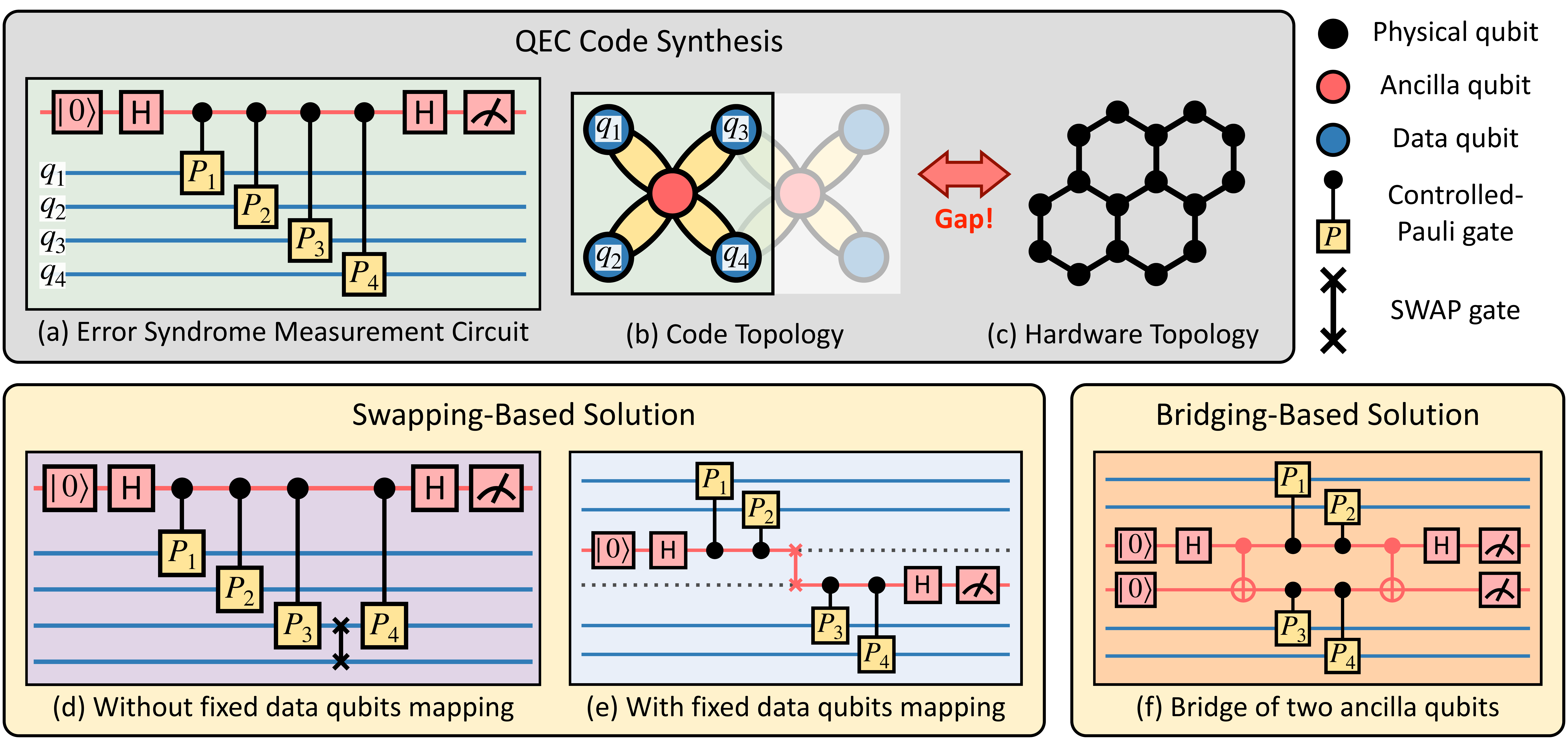}
    \vspace{-6pt}
    \caption{
    Overview of QEC code synthesis. (a) A \circuitname~circuit for error syndrome extraction. (b) The code topology associated with the \circuitname~circuits. (c) A quantum hardware with sparse architecture. (d) Implementation of swapping-based method. (e) Implementation of swapping-based method with fixed data qubits. (f) Implementation of bridging method. 
    }
    \vspace{-4pt}
    \label{fig:intro}
\end{figure*}

The core of QEC lies in the \emph{error syndrome measurement (\circuitname)} circuits constructed from the QEC codes \cite{calderbank1996good, steane1996multiple, bravyi1998quantum, bombin2007optimal, shor1995scheme, kitaev2003fault, SurfaceCode}. 
%
A typical \circuitname~circuit (\cref{fig:intro}(a)) includes an syndrome qubit (red) and multiple data qubits (blue). Controlled gates (yellow) entangle the syndrome qubit with the data qubits, allowing errors on the data qubits to affect the syndrome qubit’s state and flip its measurement outcome, indicating error occurrence. Hundreds or even more such \circuitname~circuits are executed in a QEC cycle to gather error information across the logical qubit spanning numerous data qubits. For example, a typical distance-$25$ surface code has $625$ data qubits per logical qubit and needs $624$ such \circuitname~circuits in a QEC cycle. This QEC cycle repeats during the runtime to safeguard the quantum system against errors.

However, implementing these circuits on current hardware presents a great challenge due to the mismatch between code topology and hardware architecture \cite{Surf-Stitch,chamberland2020topological, Meas_no_extra, tomita2014low, aliferis2007subsystem}. 
For instance, surface codes \cite{SurfaceCode}, a well-studied QEC code, requires a qubit connectivity of degree $4$ (\cref{fig:intro}(b)) to enable interactions between the syndrome qubit (red) and four data qubits (blue) in their \circuitname~circuits (\cref{fig:intro}(a)).
In contrast, current hardware (\cref{fig:intro}(c)) typically has sparser architectures with connectivity degrees of $2$ or $3$ \cite{gambetta2020ibm, moses2023race, Sycamore, google2023suppressing, sete2021floating, stehlik2021tunable} due to technical constraints \cite{brink2018device, malekakhlagh2020first, hertzberg2021laser}. Given the variety of QEC codes and hardware architecture, this gap is expected to persist in the foreseeable future.


A natural approach to tackle this issue involves \emph{swapping-based} methods used in qubit mapping and routing (QMR) problems~\cite{Sabre, OLSQ, molavi2022qubit}. They insert \texttt{SWAP} gates into circuits to route distant qubits adjacent for interaction. For example, in \cref{fig:intro}(a), to execute a controlled-$P_4$ gate between the syndrome qubit and distant data qubit $q_4$, a \texttt{SWAP} gate is used to relocate $q_4$ to $q_3$’s position. However, altering qubit locations with these methods is unsuitable for \circuitname~circuits because the QEC process requires consistent data qubit positions for repeated execution \cite{gottesman1998theory}. A potential solution is to enforce fixed data qubit locations as in \cref{fig:intro}(e). However, achieving this reliably with current methods may require exponentially increasing attempts as QEC circuit complexity grows.
Beyond swapping-based methods, theoretical research \cite{Flag-Bridge, chamberland2020topological, chao2018quantum, chamberland2018flag} introduces ancilla qubits in circuit mapping to form an ``ancilla bridge'' that connects data qubits (marked red in \cref{fig:intro}(f)), termed \emph{bridging} methods. Compared to swapping-based methods, they fix the qubit locations and use fewer CNOT gates (2 CNOTs compared to 3 CNOTs in a \texttt{SWAP} gate). However, current research primarily focuses on theoretical aspects and has made limited attempts only on manual designs targeting specific codes and architectures, mostly for small-scale problems. For instance, \cite{Flag-Bridge} manually implement the 7-qubit Steane code on a square lattice, while heuristic methods \cite{Surf-Stitch, chamberland2020topological} adapt the surface code \cite{SurfaceCode} to regular lattices like heavy-hexagon \cite{chamberland2020topological} up to distance 5. These specific manual designs lack adaptability to newer QEC codes like qLDPC codes \cite{tillich2013quantum, krishna2021fault} and future hardware architectures. Moreover, these methods assume ideal hardware structures and struggle with variations caused by defective qubits \cite{stace2009thresholds, stace2010error, auger2017fault}, an issue gaining recent attention \cite{lin2024codesign, suzuki2022q3de}.

We identified a crucial oversight in previous bridging methods: neglecting the broad design space enabled by ancilla qubits. Theoretically, bridging methods allow for diverse ancilla bridge shapes and sizes, various ancilla sharing designs, and flexible gate scheduling (\cref{sec:motivating example}). However, current methods rely on heuristic and manual designs, missing out on these flexibilities. This severely restricts the range of potential solutions, essentially limiting their generality.

Based on this insight, we introduce \frameworkname, the first automated compiler for mapping \circuitname~circuits using bridging methods applicable to diverse QEC codes and hardware architectures. To achieve this, we for the first time systematically categorize the flexibility of bridging methods into two key aspects: (1) ancilla bridge and (2) circuit optimization (\cref{sec:motivating example}). We then identify two critical metrics for selecting optimal solutions related to these flexibilities: (1) extra CNOT gates in ancilla bridge construction and (2) total circuit depth. Minimizing these metrics is crucial for reducing physical errors and decoherence, thereby improving circuit fidelity. Therefore, we formalize the \circuitname~circuit mapping into two optimization problems: \emph{code topology mapping} and \emph{gate scheduling} (\cref{subsec: code topology map,subsec: gate scheduling}), targeting the above two metrics. By translating code structures, hardware architectures, and mapping/scheduling conditions into Boolean constraints, we frame these optimization problems into MaxSAT framework, solvable by SAT solvers \cite{MaxSAT}. Importantly, our unified approach accommodates diverse codes and hardware setups, including systems with defective qubits. Furthermore, to handle large-scale problems, we introduce a heuristic algorithm that partitions them into manageable sub-problems addressed by our MaxSAT framework (\cref{subsec:heuristic}).

We evaluate our method against three baselines: (1) Sabre (heuristic, swapping-based) \cite{Sabre}, (2) \qmrname~(solver-based, swapping-based) \cite{molavi2022qubit}, and (3) \surfname~(heuristic, bridging-based, targeting surface codes) \cite{Surf-Stitch}. Our method achieves an average reduction of 74.9\% and 26.5\% in extra CNOT gate counts, and 34.7\% and 55.5\% in circuit depth compared to \sabrename~and \qmrname, respectively. Our method also outperforms the surface-code-specific approach \surfname~with 10\% fewer extra CNOT gates and 38.5\% shallower circuits. Additionally, these baselines frequently fail to provide solutions for special QEC codes and hardware setups and exceed memory limits for large-scale problems (\cref{sec:evaluation}). In contrast, our approach consistently identifies viable solutions. 


Overall, we make the following contributions:
\begin{itemize}
    \item We introduce \frameworkname, a bridging-based compiler for efficiently mapping \circuitname~circuits, supporting diverse QEC codes and hardware architectures.
    \item We pioneer the exploration of the expansive design space of bridging methods and formalize \circuitname~circuit mapping as a two-stage MaxSAT problem.
    \item We propose a heuristic approach based on our MaxSAT framework to effectively handle scalability challenges with large-scale codes.
    \item Our evaluation shows that \frameworkname~outperforms existing swapping-based and bridging-based methods across various QEC codes and hardware setups, while also demonstrating good scalability.
\end{itemize}

%% file: 03_background.tex
\section{Background}
\label{sec:background}
In this section, we provide essential background information on \circuitname~circuits implementation.

\subsection{\circuitname~Circuits for Stabilizer Codes}
We focus on \emph{stabilizer codes}~\cite{QCQI, djordjevic2021quantum} that encompass most extensively studied QEC codes~\cite{knill1997theory, steane1999quantum, SurfaceCode}. 

\begin{figure}[!ht]
    \centering
    \includegraphics[width=0.47\textwidth]{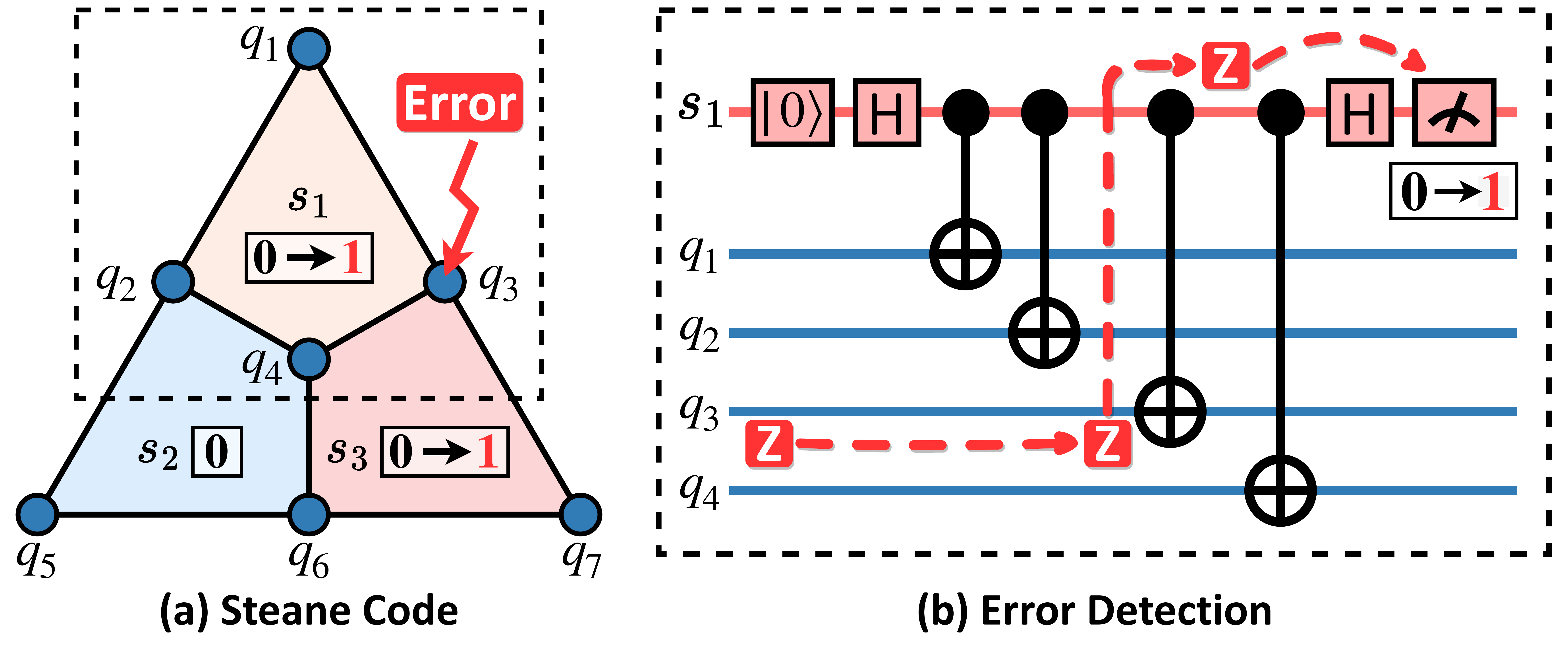}
    \vspace{-4pt}
    \caption{(a) The stabilizer structure and error syndromes of the Steane code \cite{steane1996multiple}. (b) \circuitname~circuit for the stabilizer $X_1X_2X_3X_4$ with a single syndrome qubit. A $Z$-error on $q_3$ will flip the measurement result of the syndrome qubit.}
    \label{fig:qec basic}
    \vspace{-8pt}
\end{figure}

\vspace{2pt}
\noindent\textbf{Stabilizer codes.} Stabilizer codes are characterized by a set of \emph{stabilizers}, which are Pauli strings of the form $P_1 P_2 \dots P_m$, where $P_i \in \{I, X, Y, Z\}$ denotes a Pauli operator acting on qubit $q_i$. For instance, Fig. \ref{fig:qec basic}(a) gives a stabilizer code with stabilizers $\{s_1 = X_1X_2X_3X_4, s_2 = X_2X_4X_5X_6, s_3 = X_3X_4X_6X_7\}$. Each stabilizer collects error information on its involved qubits, and combining this information from all stabilizers helps to deduce the type and location of errors, facilitating appropriate corrections. In practice, this process of error information collection involves implementing a dedicated \emph{error syndrome extraction (\circuitname)} circuit for each stabilizer.

\vspace{2pt}
\noindent \textbf{\circuitname~circuits.} The \circuitname~circuit of a stabilizer comprises a syndrome qubit (red) and several data qubits (blue), connected by controlled gates specified by the stabilizer. 
The syndrome qubit is initially set to the $|0\rangle$ state and is measured at the end. If there are no errors on the data qubits, the measurement outcome is still $0$. 
Conversely, certain errors on the data qubits will propagate through the controlled gate to the syndrome qubit, flipping its measurement outcome from $0$ to $1$, indicating error occurrence. For instance, as shown by the red dashed line in \cref{fig:qec basic}(b), a $Z$-error on qubit $q_3$ propagates through the \texttt{CNOT} gate and flipsthe measurement outcome of $s_1$.
Notably, stabilizers often share data qubits, which introduces complexity to the mapping and scheduling of their \circuitname~circuits (more details in \cref{sec:motivating example}).

\vspace{2pt}
\noindent\textbf{Error detection and correction.} During a QEC round, each \circuitname~circuit is executed once. Their measurement results consist \emph{error syndromes} that guide error correction. For instance, in \cref{fig:qec basic}(a), the error syndrome $(s_1, s_2, s_3) = (1,0,1)$ suggests a $Z$-error on qubit $q_3$, which can be corrected by applying a $Z$ gate to $q_3$ since $Z^2 = I$. Crucially, all \circuitname~circuits can theoretically operate in parallel within a QEC round due to the commutativity of stabilizers, allowing for a larger optimization space for scheduling (more details in \cref{sec:motivating example}).

\subsection{Bridging method}
\circuitname~circuits necessitate interactions between the syndrome qubit and all data qubits, which is challenging with hardware of limited connectivity. The \emph{bridging method} \cite{Flag-Bridge} offers a promising solution by extending the syndrome qubit with additional ancilla qubits called \emph{bridge qubits}, forming an “ancilla bridge” that connects distant data qubits. This approach resolves connectivity issues while maintaining error syndrome extraction as in the original \circuitname~circuit.

\begin{figure}[!t]
    \centering
    \includegraphics[width=0.47\textwidth]{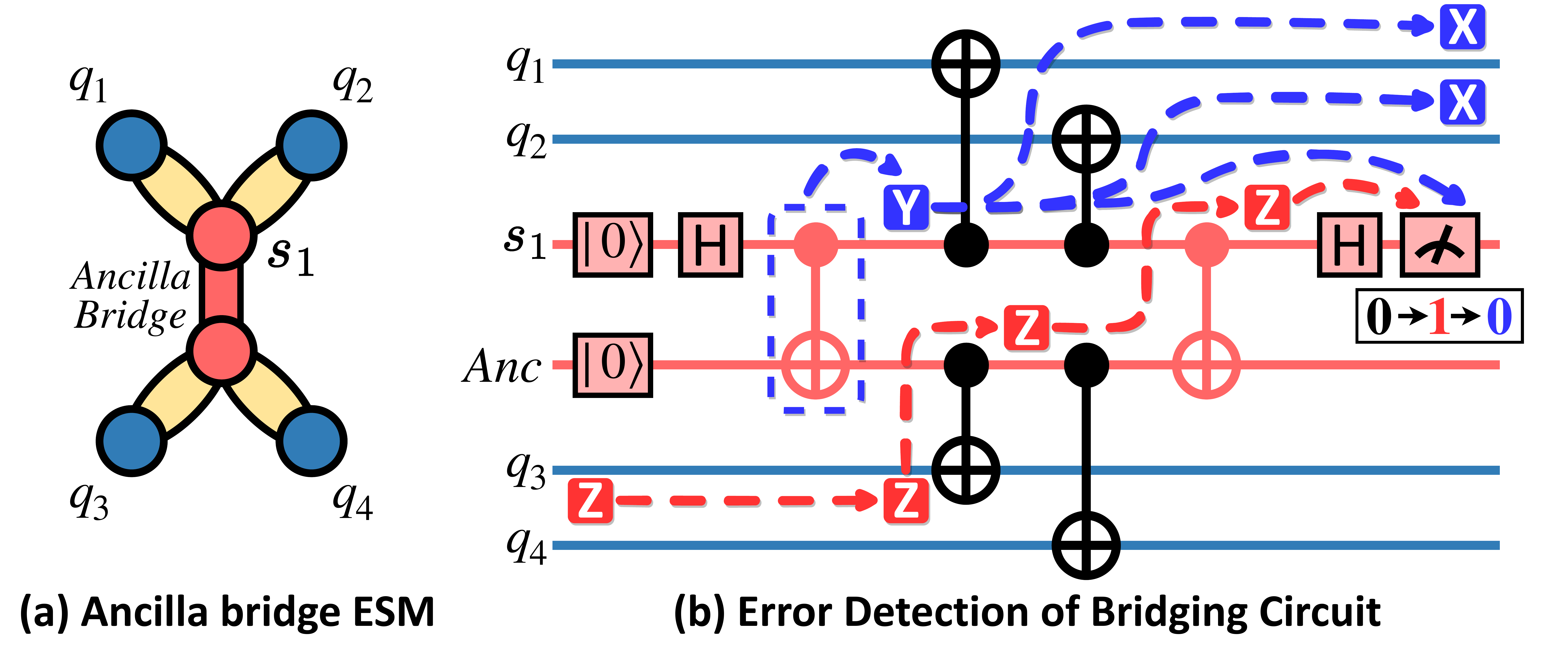}
    \vspace{-4pt}
    \caption{Bridging method. (a) The connectivity of the bridging-based \circuitname~circuit. (b) Red line: This construction maintains error detection. Blue line: The inserted \texttt{CNOT} gate may introduce more error leading to error syndromes.}
    \label{fig:F-T_syndrome_extraction} 
    \vspace{-8pt}
\end{figure}

\vspace{2pt}
\noindent\textbf{Mitigating connectivity issue.} In bridging method, the ancilla bridge is prepared in a special entangled state \cite{greenberger1989going} by inserting additional \texttt{CNOT} gates (highlighted in red). 
This state ensures that the controlled gates acting on the syndrome qubit can be transferred to the bridge qubit while maintaining circuit equivalence. 
As a result, the connectivity requirements, originally concentrated on the single syndrome qubit, are now distributed across multiple bridge qubits, thereby reducing the overall connectivity demand. 
For instance, the circuit with a length-2 ancilla bridge in \cref{fig:F-T_syndrome_extraction}(b) only requires each qubit to connect to at most 3 others, compared to 4 in \cref{fig:qec basic}(b), allowing for a direct mapping to current sparse quantum architecture~\cite{gambetta2020ibm, moses2023race, Sycamore, sete2021floating}.

\noindent\textbf{Error syndrome extraction.} Remarkably, the ancilla bridge construction ensures that errors on the data qubits trigger a flip in the measurement result of the syndrome qubit, just as in the original \circuitname~circuit (Fig. \ref{fig:qec basic}(b)). This occurs because errors can still propagate to the syndrome qubit through the special entangled state of the ancilla bridge (GHZ state). For instance, in Fig. \ref{fig:F-T_syndrome_extraction}(b), a $Z$-error on $q_3$ still flips the measurement result of the syndrome qubit $s_1$ (as shown by the red dashed line). Consequently, the transformed \circuitname~circuit with bridge qubits preserves the same error detection and correction capabilities as the original circuit.

\begin{figure*}[!ht]
    \centering
    \includegraphics[width=0.99\textwidth]{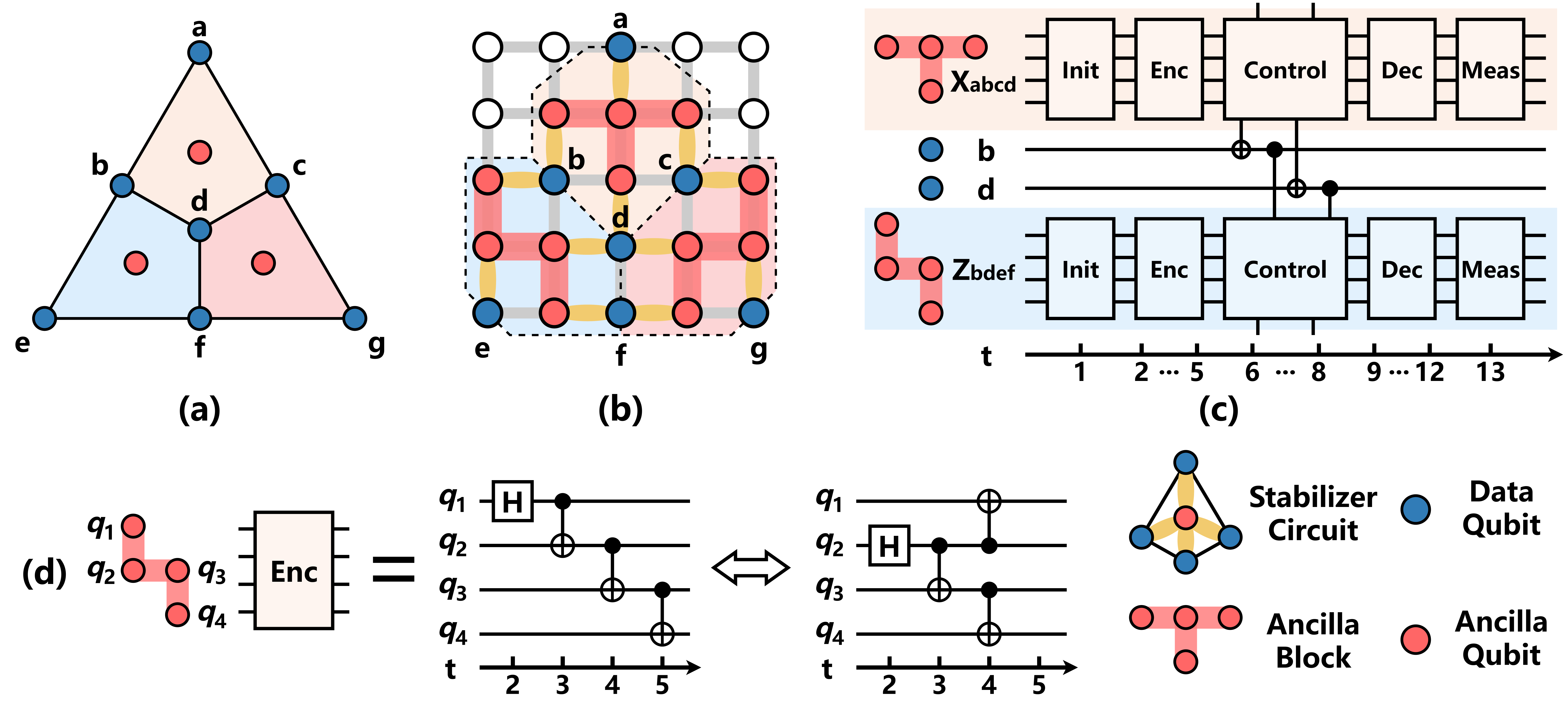}
    \caption{Flexibility of the bridging method for \circuitname~circuits synthesis. (a) Code Topology Mapping: The ancilla bridges of different stabilizer may have different size, shape and may share ancilla qubits. (b) Circuit Scheduling: The scheduling need to consider ESM commutativity, equivalent state preparation on ancilla bridges and correct circuit parallelism.}
    \label{fig:motivation}
\end{figure*}

\noindent\textbf{Impact of extra \texttt{CNOT} gates. }Bridging methods introduce additional \texttt{CNOT} gates (highlighted in red) compared to the original \circuitname~circuit (Fig. \ref{fig:qec basic}(b)). Similar to the extra \texttt{CNOT} gates in SWAP-based methods, these \texttt{CNOT}s can introduce errors to data qubits, potentially increasing the logical error rate. For example, in Fig. \ref{fig:F-T_syndrome_extraction}(b), if a $Y$-error occurs due to an introduced \texttt{CNOT} gate (within the block), it can induce a correlated two-qubit $XX$ error on data qubits and affect the accuracy of the syndrome measurement. Therefore, minimizing the number of \texttt{CNOT} gates is crucial to mitigate these detrimental effects.

%% file: 04_formulation.tex
\section{Motivating Examples}\label{sec:motivating example}
In this section, we provide examples to illustrate the vast design space of bridging-based \circuitname~circuit mapping and scheduling. We highlight two main flexibilities: (1) varying ancilla bridges to optimize qubit mapping with fewer \texttt{CNOT} gates, and (2) leveraging circuit commutativity and equivalence to optimize gate scheduling for reduced circuit depth.





\subsection{Flexibility of Code Topology Mapping}
Ancilla bridges offer significant flexibility, particularly in terms of \emph{(1) various shapes and sizes} and \emph{(2)  shared ancilla qubits}. This flexibility helps to find an effective mapping of the stabilizer code's topology onto the architecture, including the data qubit mapping and the ancilla bridge allocation. Moreover, varying constructions can introduce different numbers of \texttt{CNOT} gates (\cref{sec:background}), impacting the fidelity of the synthesized \circuitname~circuits. Since the total size of the ancilla bridge linearly correlates with the number of extra \texttt{CNOT} gates, we prioritize minimizing it as our optimization goal in code topology mapping (\cref{sec:technical}).

\vspace{2pt}
\noindent\textbf{(1) Various shapes and sizes.} 
The flexibility of ancilla bridges extends to their shapes and sizes, which stems from their operational mechanism. 
As long as the ancilla bridge is initialized in an entangled state, errors on data qubits can propagate to the syndrome qubit regardless of which bridge qubit they are connected to. Thus, the circuit can extract the error syndrome, irrespective of the ancilla bridge's shape or size. Consequently, the shape and size of the ancilla bridge can vary depending on the allocation of data qubits. For example, as shown in \cref{fig:motivation}(a), both the 4-qubit and 5-qubit ancilla bridges for $s_1$ and $s_3$ can effectively fulfill their function, demonstrating this flexibility. For more details, we recommend referring to Lao et al. \cite{Flag-Bridge}.



\vspace{1pt}
\noindent\textbf{(2) Ancilla qubit sharing.} Ancilla bridges for different stabilizers can share the same ancilla qubits, reducing the number of ancilla qubits required to be connected to a data qubit. For example, in \cref{fig:motivation}(a), stabilizers $s_1$ and $s_2$ share two ancilla qubits in the middle row. Although this design saves qubits, \circuitname~circuits sharing ancilla qubits cannot be executed simultaneously due to resource conflicts. This presents a challenge for gate scheduling to optimize parallel executions.



\subsection{Flexibility of \circuitname~Circuit Scheduling}
The distinctive structures of \circuitname~circuits allow for more flexible optimizations in scheduling compared to the gate-level scheduling for general quantum circuits. This flexibility arises mainly from three aspects: \emph{(1) commutativity of stabilizer measurements}, \emph{(2) equivalent state preparations on ancilla bridges} and \emph{(3) parallel \circuitname~circuits}. Leveraging this flexibility, our goal is to identify a correct scheduling while minimizing the circuit depth.

\vspace{2pt}
\noindent\textbf{(1) Commutable stabilizer measurements.} Implementing a QEC code involves executing \circuitname~circuit blocks (\cref{fig:qec basic}(b)) for all stabilizers. Theoretically, these \circuitname~circuit blocks for various stabilizers commute, allowing for arbitrary permutations in their order. 
As illustrated in \cref{fig:motivation}(b1), sequentially executing stabilizer blocks $s_1$, $s_2$, and $s_3$ can be optimized by moving $s_3$ to the front and executing it in parallel with $s_1$, leading to a substantial decrease in circuit depth.

\vspace{2pt}
\noindent\textbf{(2) Equivalent state preparations on ancilla bridges.} As explained in \cref{sec:background}, the ancilla bridge is initialized in an appropriate entangled state. This initialization can be accomplished through various circuits and some of which may offer smaller circuit depths \cite{greenberger1989going}. 
For instance, in \cref{fig:motivation}(b2), both circuits can prepare the 4-qubit initial state on the ancilla bridge of $s_1$, but the second one has a smaller circuit depth.

\vspace{2pt}
\noindent\textbf{(3) Parallel \circuitname~circuits.} Although two \circuitname~circuits can be executed in parallel as long as their ancilla bridges do not overlap, the overlap on data qubits makes it impossible to arbitrarily order the controlled gates while still ensuring a correct error syndrome extraction. To illustrate, as depicted in \cref{fig:motivation}(b3), the simultaneous measurement of two stabilizers $Z_1Z_2$ and $X_1X_2$, which share two data qubits, poses a challenge for the scheduling of their controlled-$Z$ and controlled-$X$ gates. While the second circuit provides correct measurement results, the first one can only yield random outcomes \cite{SurfaceCode}. This necessity for correctness adds intricacy to scheduling and must be addressed accordingly.


\section{Overview of \frameworkname}
\label{sec:overview}
This section outlines the workflow of our \frameworkname~framework. It splits into two paths based on problem size, as shown in \cref{fig:overview}. For small-scale problems, \frameworkname~utilizes our dedicated two-stage SAT solver (\cref{fig:overview}, left frame) to generate optimal \circuitname~circuit implementations. For large-scale problems, we employ the relaxation technique of \emph{problem partitioning} to resolve the scalability issue (\cref{fig:overview}, right).

\subsection{Constraint-based Approach: Two-stage SAT}
\label{subsec optimal solver}
The left frame in \cref{fig:overview} represents the central component of the \frameworkname~framework. It takes the stabilizer code and hardware architecture as inputs and generates the \circuitname~circuit implementation, specifying both \emph{qubit allocation} and \emph{scheduled circuit}. Building on the two flexibility categories identified in \cref{sec:motivating example}, we break down this problem into two stages of constraint-based optimization: \emph{Stage 1. Code topology mapping} and \emph{Stage 2. \circuitname~circuit scheduling}.

\vspace{2pt}
\noindent\textbf{Stage 1. Code Topology Mapping.}
This stage aims to map data qubits and allocate ancilla bridges based on the stabilizer code and hardware architecture. The code is represented by its stabilizer set $Stab = \{s_1,s_2,\cdots\}$, where $s_i$ are Pauli strings acting on data qubits $Data = \{q_1,q_2,\cdots\}$, while the hardware architecture is represented by its coupling graph, with vertices being physical qubits $Phys = \{P_1,P_2\cdots\}$ and the edges $Edges = \{e_1,e_2,\cdots\}$ specifying their connectivity. The output consists of: \emph{(1) Data qubit mapping} $\Pi:Data\rightarrow Phys$ that map data qubits to physical qubits on hardware, \emph{(2) Ancilla bridge }$Anc: Stabs\rightarrow \text{Subsets of }Phys$  that allocate an ancilla bridge for each stabilizer, and \emph{(3) Coupling relation }$CP: Stabs\times Data\rightarrow Phys$ that specifies the ancilla qubit to which each data qubit connects in each stabilizer.

We address this problem using the MaxSAT framework. The validity of the output is ensured by various \emph{constraints} (red frame), while optimality is pursued by maximizing an objective function composed of a weighted sum involving multiple optimization goals (blue frame). For instance, we minimize the overall size of ancilla bridges, which directly impacts the number of introduced \texttt{CNOT} gates (\cref{sec:motivating example}), and maximize the number of compatible stabilizers to enable parallel execution. The corresponding objective function is:
\begin{equation*}
\begin{split}
 -w_1\cdot 
    \#\left\{
    \begin{aligned}
         & \text{Ancilla} \\
         & \text{qubits} \\
    \end{aligned}
    \right\}
    + w_2\cdot 
    \#\left\{
    \begin{aligned}
         &  \text{Compatible anc.} \\
         &  \text{\quad bridge pairs}\\
    \end{aligned}
    \right\}
\end{split}
\end{equation*}
Adjusting the weights $w_1,w_2$ alters the priority of optimization goals.
The detailed MaxSAT encoding of constraints and objective function will be elaborated in \cref{sec:technical}.

\begin{figure}[!t]
    \centering
    \includegraphics[width=0.47\textwidth]{Figures/Overview.pdf}
    \caption{Overview of \frameworkname.}
    \label{fig:overview} 
    \vspace{-10pt}
\end{figure}

\vspace{2pt}
\noindent\textbf{Stage 2. Gate Scheduling.}
This stage aims to find a gate scheduling with minimal circuit depth, based on the qubit allocation determined in Stage 1. The output is a mapping $Sche: Circuits \rightarrow \{1,2,\cdots\}$ that assigns each gate $g\in Circuits$ a time step for execution. The search process is encoded into a SAT solver, with constraints (detailed in \cref{sec:technical}) specified to ensure scheduling validity. To minimize circuit depth, we iteratively query the SAT solver with decreasing requirements on circuit depth until no solutions are found. This iterative approach helps obtain increasingly better solutions with smaller circuit depths, finally leading to a solution with minimal depth.

The outputs of the above two stages together constitute the complete output of our framework, specifying the mapping of data qubits and ancilla bridges, as well as the time schedule of each gate in the \circuitname~circuit.


\subsection{Heuristic Approach: Code Partitioning}
The optimal approach described in \cref{subsec optimal solver} faces challenges when applied to large-scale problems due to the escalating number of constraints. To address this scalability concern, we partition the stabilizer set $S$ of a code into smaller subsets $S_1,\cdots,S_k$ and leverage our two-stage optimal solver described in \cref{subsec optimal solver} to find their implementations. This yields optimized \circuitname~circuits $C_i$ for each subset $S_i$. Subsequently, we sequentially execute these \circuitname~circuits for sub-problems to generate the complete solution for the entire stabilizer code. SWAP gates are inserted between adjacent \circuitname~circuits to reposition the qubits from the previous circuit $C_i$ to meet the requirements of the next one $C_{i+1}$ (detailed in \cref{sec:technical}). The evaluation (\cref{sec:evaluation}) shows that our approach enables scalability while yielding favorable results.

%% file: 05_tech.tex
\section{Constraint-Based Code Synthesis}
\label{sec:technical}
In this section, we delve into the technical aspects of SAT encoding for the two stages in \frameworkname: \emph{code topology mapping} and \emph{\circuitname~circuit scheduling}.



\subsection{Stage 1: Code Topology Mapping via MaxSAT}
\label{subsec: code topology map}
In this stage, we encode the mapping of data qubits and the allocation of ancilla bridges into a MaxSAT problem. We formulate the qubit location requirements as \emph{hard constraints} and the objectives of minimizing the total ancilla bridge size and incompatible stabilizer count as \emph{soft constraints}.

\subsubsection{Hard Constraints:}
We list a few hard constraints that ensures the validity of qubit allocation.

\vspace{2pt}
\noindent \textbf{Hard A. Basic Requirements.} A legitimate qubit allocation should satisfy the following basic requirements.\\
(1) Each data qubit is mapped to a unique physical qubit: 
\begin{align*}
\sum_{p \in Phys} map(q, p) = 1 \text{,~for~} q \in Data \\
\sum_{q \in Data} map(q, p) \leq 1 \text{,~for~} p \in Phys
\end{align*}
where $map(q,p)=True$ if data qubit $q$ is mapped to physical qubit $p$, and $map(q,p)=False$ otherwise.\\
(2) A physical qubit cannot serve as both a data qubit and an ancilla qubit simultaneously:
\begin{align*}
    \neg map(q, p) \bigvee \neg anc(s, p) \text{,~for~} (q,~s) \in Data \times Stabs
\end{align*}
where $anc(s,p)=True$ if physical qubit $p$ is an ancilla qubit for stabilizer $s$, and $anc(s,p)=False$ otherwise.





\vspace{2pt}
\noindent \textbf{\textit{Hard B.} Connectivity of ancilla bridge.} The ancilla bridge for each stabilizer needs to form a connected graph. To ensure this, we employ breadth-first traversal (BFT) \cite{BFT}. All physical qubits in the connected $Graph[Anc[s]]$ must be visited within a specified parameter $L[s]$, representing the maximum size of the ancilla bridge $Anc[s]$ for stabilizer $s$. We introduce auxiliary variables $v_{s,p}(p_i,t)$ to track whether physical qubit $p_i$ has been visited after the $t$-th round ($t \in \{1, \dots, L[s]\}$). In the initial round, we ensure that the starting vertex is visited:
\begin{align*}
v_{s, p}(p, 1) \bigwedge_{p_i \neq p} \neg v_{s, p}(p_i, 1)
\end{align*}
In subsequent rounds, an unvisited physical qubit $p_i$ in the ancilla bridge $Anc[s]$ is visited if any of its adjacent qubits $p_j$ have been visited in the previous rounds:
\begin{align*}
v_{s, p}(p_i, t) &\rightarrow~ anc(s, p_i) \\
&\wedge \left[ v_{s, p}(p_i, t-1) \bigvee_{p_j \in Adj(p_i)} v_{s, p}(p_j, t-1) \right]
\end{align*}
To ensure that the traversal starts from any possible physical qubit, we enforce constraints for each $p \in Phys$:
\begin{align*}
anc(s, p) \rightarrow \bigwedge_{p_i \in Phys} (\neg anc(s, p_i) \vee v_{s, p}(p_i, L[s]))
\end{align*}
The number of variables and the conjunctive normal form (cnf) clause used for each stabilizer are both $O(L[s] * |Stabs| ^ 2)$.

\vspace{2pt}
\noindent \textbf{\textit{Hard C.} Coupling between data and ancilla qubits.}
Each data qubit for stabilizer $s$ must be directly coupled to some ancilla qubit in its ancilla bridge $Anc[s]$:\\
\noindent (1) Exactly one ancilla qubit $p$ is coupled with data qubit $q$:
\begin{align*}
    \sum_{p \in Phys} cp(s, q, p) = 1 \text{,~for~} q \in Data[s]
\end{align*}

\noindent (2) The coupling relation $CP(s, q) = p_i$ requires $q$ to be mapped to some qubit $p_i$ adjacent to $p$:
\begin{align*}
    cp(s, q, p) \rightarrow \bigvee_{p_i \in Adj(p)} map(q, p_i) \bigwedge anc(s, p)
\end{align*}

\subsubsection{Soft Constraints:}
We introduce several soft constraints to optimize the performance of \circuitname~circuit implementation. Each soft clause is assigned a non-negative weight indicating its priority, and a MaxSAT solver is used to obtain a solution that maximizes the sum of the weights of all satisfied soft clauses.

\vspace{2pt}
\noindent \textbf{\textit{Soft P1.} Minimizing the total ancilla bridge size.} 
Reducing the total ancilla bridge size compresses the \circuitname~circuit and minimizes the introduced extra CNOT gates, enhancing circuit fidelity. This is achieved by maximizing the number of Boolean variables $anc(s, p)$ set to $False$. Hence, we add the following soft constraints with weight $P1$:
\begin{align*}
    \neg anc(s, p) \text{,~for~} (s, p) \in Stabs \times Phys
\end{align*}

\vspace{2pt}
\noindent \textbf{\textit{Soft P2.} Mitigating the stabilizer conflicts.} 
Stabilizers without conflicts in ancilla qubits can be efficiently executed in parallel, termed as \emph{compatible stabilizers}. Our aim is to maximize the count of compatible ancilla bridge pairs. Thus, we introduce the following soft constraints with weight $P2$:
\begin{align*}
    \bigwedge_{p \in Phys} \neg anc(s, p) \vee \neg anc(s', p) \text{, for } s, s' \in Stabs
\end{align*}

\cref{fig:Soft} illustrates how different optimization objectives impact the mappings of the same code $Stabs = {X_1X_2X_3, Z_2Z_3}$ onto a heavy square architecture. In the first row, Soft P1 is the primary objective, with weight $P_1$ notably greater than $P_2$. Consequently, the resulting mapping features an ancilla bridge with size $1$, yet two stabilizers share the same ancilla qubit. Conversely, in the second row, Soft P1 is prioritized, resulting in a mapping without conflicts in data qubits, albeit with a larger ancilla bridge size. In practice (\cref{sec:evaluation}), we prioritize Soft P1 over Soft P2, given its more significant impact on the overall error correction performance.

\begin{figure}[!h]
    \centering
    \includegraphics[width=0.47\textwidth]{Figures/Influence1.pdf}
    \caption{Different mappings for stabilizer code $\{s, s'\}$ with different optimization priority.}
    \label{fig:Soft}   
\end{figure}




\subsection{Stage2: \circuitname~Circuit Scheduling via SAT}
\label{subsec: gate scheduling}
This stage encodes the gate scheduling of \circuitname~circuits into a SAT problem. We introduce additional Boolean variables to characterize all operations within the circuits (\cref{subsec: identify operation}). These variables help to formulate constraints based on dependencies between operations, ensuring the validity of their scheduling (\cref{subsec. boolean constraints}). Finally, we iteratively query the SAT solver to achieve a scheduling with minimal circuit depth.

\begin{figure}[!ht]
    \centering
    \vspace{2pt}
    \includegraphics[width=0.47\textwidth]{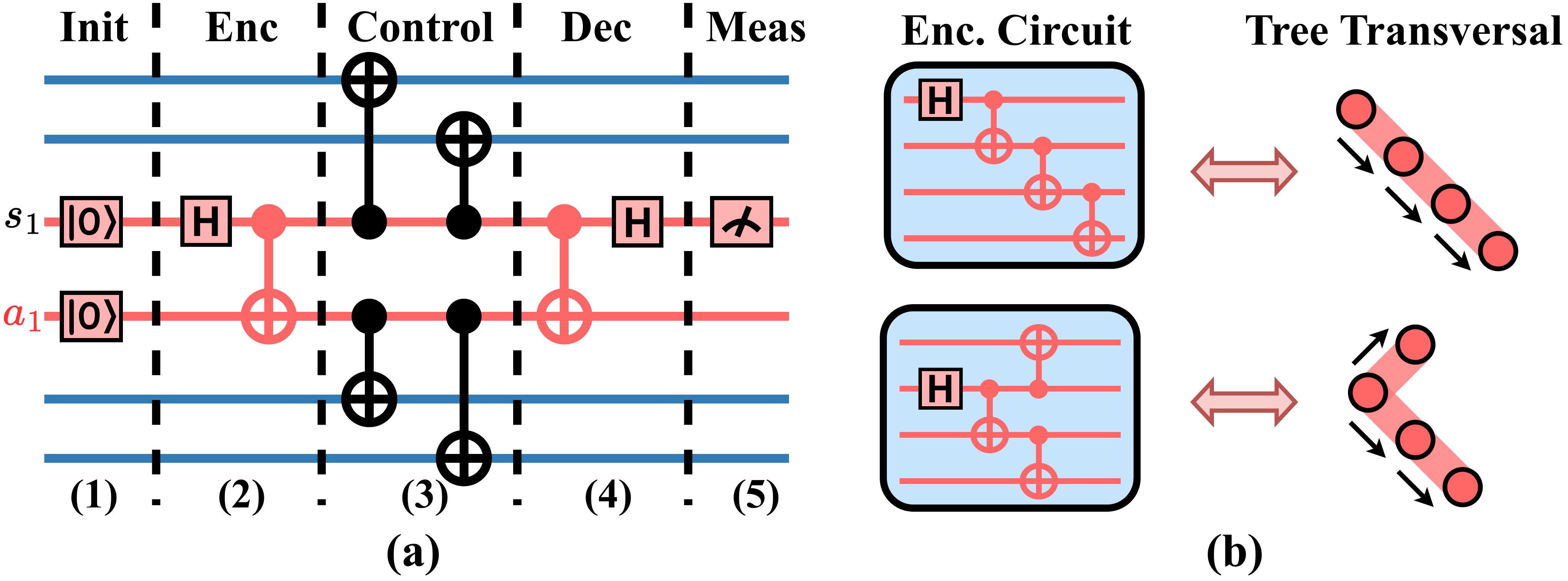}
    \vspace{-6pt}
    \caption{(a) Five stages of a typical \circuitname~circuit with ancilla bridge. (b) Equivalence between encoding (decoding) circuit for ancilla bridge and tree transversal.}
    \label{fig:fivestage}   
    \vspace{-6pt}
\end{figure}

\subsubsection{Operations within \circuitname~Circuits:}
\label{subsec: identify operation}
A \circuitname~circuit with ancilla bridge typically comprises five phases: \emph{(1) \textit{\phaseonename}}, \emph{(2) \textit{\phasetwoname}}, \emph{(3) \textit{\phasethreename}}, \emph{(4) \textit{\phasefourname}}, and \emph{(5) \textit{\phasefivename}}, as shown in \cref{fig:fivestage}(a). Operations in Phases (1)(3)(5) are determined by the output of Stage 1, which includes qubit allocation and coupling relations between ancilla and data qubits. Specifically, qubit allocation specifies all ancilla qubits initialized in the $|0\rangle$ state, with the syndrome qubit measured at the end, constituting Phases (1)(5). The coupling relation precisely determines all CNOT gates in Phase (3). To characterize Phases (2)(4), we introduce the following Boolean variables:

\noindent \ballnumber{1} $enc(s, p, p’)$ and $dec(s, p, p’)$:
The variable $enc(s, p, p’)$ is set to $True$ if there is a $CNOT$ gate from $p$ to $p’$ in the \textit{\phasetwoname} phase for stabilizer $s$, and $False$ otherwise. Similar definitions apply to $dec(s, p, p’)$. Additionally, a Hadamard gate is applied to the syndrome qubit. These Boolean variables collectively characterize the operations in Phases (2)(4).

\vspace{2pt}
\noindent \ballnumber{2} $time(g, t)$:
This denotes the time assignment of operations $g\in Circuits$. If the operation $g$ is executed at time $t$, $time(g, t)$ is set to $True$; otherwise, it is set to $False$.

\subsubsection{Boolean Constraints for Gate Scheduling:}
\label{subsec. boolean constraints}
This section outlines the constraints that must be satisfied by a valid gate scheduling, using the variables defined above.


\vspace{2pt}
\noindent \textbf{\textit{Hard A. Time Assignment Constraints.}}
A valid gate scheduling should satisfy the following basic requirements regarding time assignments. We define $T$ as the maximum allowable circuit depth for the search.\\
\noindent (1) All gates should be executed exactly once:
\begin{align*}
    \sum_{t} time(g, t) = 1 \text{, for } g \in Circuits
\end{align*}

\noindent (2) Two operations $g$ and $g'$ act on the same physical qubit $p$ cannot be executed simultaneously. In other words, at most one operation acting on $p$ can be executed at time $t$:
\begin{align*}
    \sum_{g \in Act[p]} time(g, t) \leq 1 \text{, for } t \in \{1, ..., T\}
\end{align*}
where $Act[p]$ denotes all the operations acting on $p$. 




\vspace{2pt}
\noindent \textbf{\textit{Hard B.} Valid $Enc[s]$ and $Dec[s]$ circuits.} 
The circuits in the \textit{\phasetwoname} and \textit{\phasefourname} phases should effectively encode and decode the GHZ state for the ancilla bridge. We observe that the encoding circuit’s formulation is analogous to a tree traversal, where the syndrome qubit serves as the root with an H gate acting on it, and other ancilla qubits are visited via CNOT links (\cref{fig:fivestage}(b)). The decoding circuit can be viewed as the mirror image of the encoding circuit. With this understanding, we establish the following constraints:

\noindent (1) The transversal tree must have exactly one root:
\begin{align*} 
    \sum_{p \in Anc[s]} enc(s, p, p) = 1
\end{align*}
where $enc(s, p, p)$ is set to $True$ if an H gate acts on physical qubit $p$, and $False$ otherwise.

\noindent (2) Each node within the transversal tree is visited exactly once, ensuring that every ancilla qubit $p \in Anc[s]$ is subjected to exactly one $CNOT$ gate targeting it:
\begin{align*} 
    \sum_{p_i \in Anc[s]} enc(s, p_i, p) = 1
\end{align*}

\noindent (3) The control qubit must be prepared before its associated CNOT gate is executed. Specifically, if $g_j = CNOT(p_i, p_j)$ and $g_i = CNOT(p_k, p_i)$, then $g_j$ must be executed after $g_i$:
\begin{align*} 
    enc(s, p_i, p_j) \rightarrow \left( t_{g_i} < t_{g_j} \right) \text{, for } p_i \neq p_j \in Anc[s]
\end{align*}


\vspace{2pt}
\noindent \textbf{\textit{Hard C.} Correct order of $Ctrl[s]$ circuits}.
The anticommuting controlled-$P_i$ gates in the \textit{\phasethreename} phase should satisfy a particular order to ensure the correctness of the circuit. Specifically, the order is correct if and only if the count of anti-commuting gate pairs $(c, c')$ satisfying $t_c < t_{c'}$ is an even number. This requirement is encoded into the constraint:
\begin{align*}
    \bigoplus_{Anti(c, c')} (t_c < t_c') = 0 \text{, for } c \in Ctrl[s] \text{~~and~} c' \in Ctrl[s']
\end{align*}
where $\bigoplus$ means the mod 2 addition and $Anti(c, c')=True$ means $c$ and $c'$ anti-commute.

\vspace{2pt}
\noindent \textbf{\textit{Hard D.} Commutativity of incompatible stabilizers.} 
Two stabilizer circuits $s$ and $s'$ sharing ancilla qubits cannot he executed simultaneously.
This constrain can be realized by requiring $Init[s]$ after $Meas[s']$, $Init[s']$ after $Meas[s]$:
\begin{align*} \label{Hard C}
    \left[ \bigwedge \limits_{p \in Share} (t_{Mz'[p]} < t_{Init[p]}) \right] 
    \bigvee 
    \left[\bigwedge \limits_{p \in Share} (t_{Mz[p]} < t_{Init'[p]})\right]
\end{align*}
where $Share = Anc[s] \cap Anc[s']$ represent all physical qubit that shared by two stabilizer's ancilla block.

\subsection{Heuristic Approach}
\label{subsec:heuristic}
As previously discussed, the number of clauses increases exponentially with the number of stabilizers in a QEC code, creating a significant bottleneck in our framework. To address this challenge for large-scale QEC code synthesis, we propose a relaxation method that can still yield locally optimal solutions. Our approach involves partitioning the set of stabilizers and solving a SAT problem for each subset sequentially through our two-stage constraint-based approach. Finally, we insert \textit{routing layers} between the these synthesized \circuitname~circuits to integrate them into the original code.

\vspace{2pt}
\noindent\textbf{Partitioning the stabilizer set.} We can think of a stabilizer set $S$ as a union of subsets, $S_0 \cup \dots \cup S_k$. However, due to the commutativity of the stabilizers, there are millions of possible partition choices. Since each stabilizer can be viewed as a set of constraints for the data qubits to satisfy, we want the constraints for the same data qubits to be in the same partition, allowing them to be handled together. It means to minimize the connections or shared qubits between stabilizers from different subsets. 
As a result, we formulate this as a simple balanced graph partitioning problem in graph theory \cite{andreev2004balanced}. 
For example, in the Steane code shown in \cref{fig:qec basic}(a), we treat each stabilizer as a vertex. As stabilizer $s_1$ shares two data qubits, $q_2$ and $q_4$, with stabilizer $s_2$, two edges will be added to the graph. The balanced graph partitioning divides the vertices into $k$ components of almost equal size, minimizing the capacity of edges between different components.

\vspace{2pt}
\noindent\textbf{Sorting the SAT problems.} After partitioning the stabilizer set, we can apply our two-stage SAT solver to synthesize the \circuitname~circuit for each stabilizer subset. However, different resulting data qubit mappings can make it challenging to integrate them into a complete circuit for the original code. 
The mapping $map_i$ for subset $S_i$ should follow the mappings from $map_0$ to $map_{i-1}$ as closely as possible, because stabilizers from different subsets may share the same data qubits, and we want these qubits to be mapped to the same positions in $map_i$.
Therefore, after solving the SAT problem for $S_1, \dots, S_i$, we select the subset $S_j$ that shares the most data qubits with all data qubits involved in $S_1 \cup \dots \cup S_i$ as the next SAT problem to solve. Moreover, we add soft constraints in Stage 1 to maximize the number of shared data qubits that remain in their previous positions:
\[ \max \left| \{ q \mid map_i(q, p) \vee map_{0 \dots i}(q, p) \} \right| \]

\vspace{2pt}
\noindent\textbf{Integrating the \circuitname~circuits.} After all \circuitname~circuits for each subset are generated, we connect them to form the complete \circuitname~circuit for the original QEC code. Although the circuits may require different data qubit mappings and ancilla bridge allocations on the architecture, we insert integration layers between them to move the data qubits to meet these varying requirements. Since all ancilla qubits function identically and only data qubits contain the logical information, we only need to route the data qubits instead of all used physical qubits, as general QMR methods do.
In the integration layer, to connect \circuitname~circuits, we find the shortest path for each data qubit that changes its position and insert \textit{SWAP} gates to move it to the new position.

%% file: 06_evaluation.tex
\section{IMPLEMENTATION AND EVALUATION}
\label{sec:evaluation}

\subsection{Experimental Setup}
\label{sec:setup}
\noindent \textbf{Evaluation setting:} 
Our framework, \frameworkname, utilizes the state-of-the-art MaxSAT solver, NuWLS-c \cite{NuWLS-c}, and the widely-used SAT solver, PySAT \cite{PySAT}, with a solving time limit of 7,200 seconds and a memory limit of 128 GB. For logical error analysis, we used Stim \cite{stim}, a widely-used QEC simulator for \circuitname~circuits, and PyMatching \cite{higgott2022pymatching}, a error syndrome decoder that implements the Minimum Weight Perfect Matching (MWPM) algorithm.

\begin{figure}[!h]
    \centering
    \includegraphics[width=0.45\textwidth]{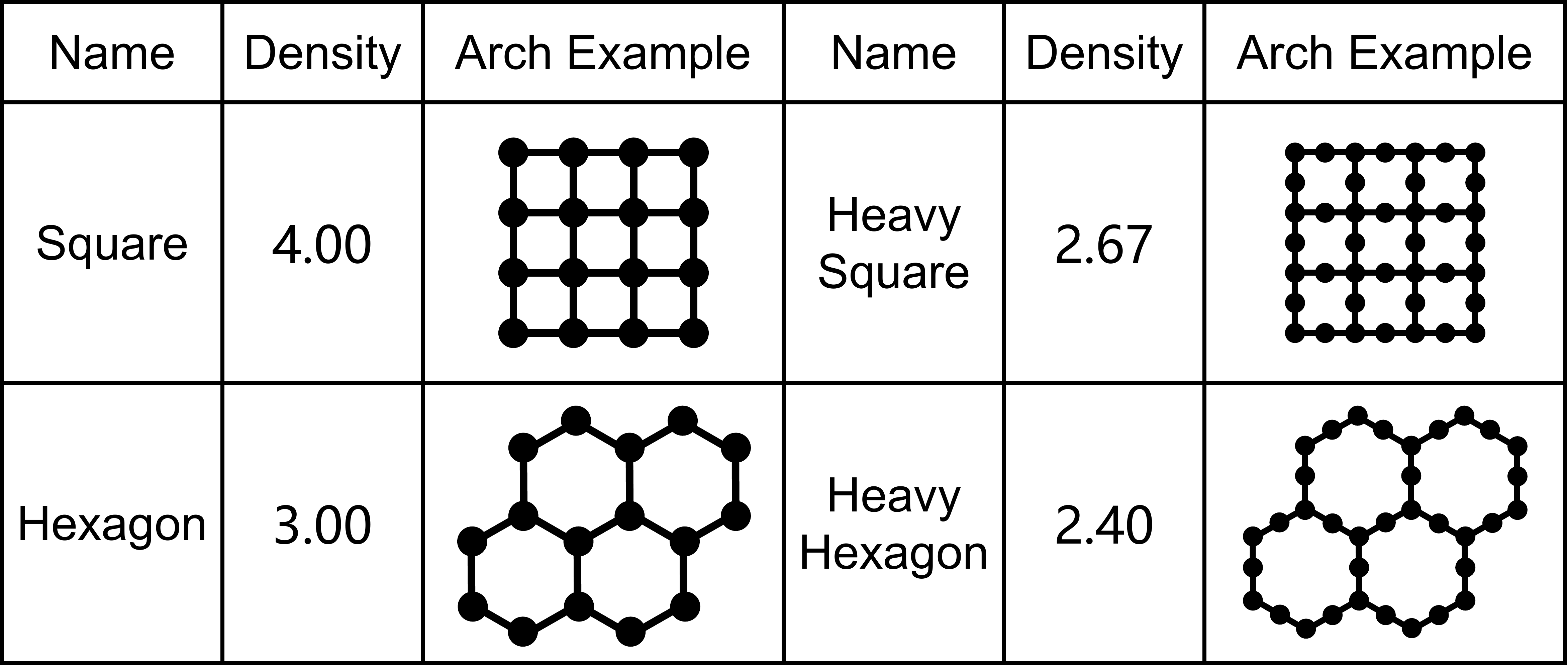}
    \caption{Overview of device architectures.}
    \label{fig:CG}
    \vspace{-8pt}
\end{figure}

\vspace{2pt}
\noindent \textbf{Hardware architectures:} 
We cover a diverse range of practical device architectures in our evaluation. As shown in \cref{fig:CG}, the selected architectures are sourced from the latest quantum machines, including Google's Sycamore \cite{Sycamore}, IBM's Osprey, and IBM Q Tokyo \cite{chow2021ibm}.
To represent the connectivity of the architecture, we use the average degree of the nodes, denoted as $Density(G) = 2 |Edge| / |Node|$. 

\begin{table*}[ht]
    \centering
    \caption{Results of \frameworkname~and the baselines for various stabilizer codes across different device architectures.}
    \resizebox{0.95\textwidth}{!}{
        \begin{tabular}{|c|c|c|c||*{3}{c|}| *{3}{c|}| } 
        
        \hline
        \multirow{2}{*}{Code} & \multirow{2}{*}{Density} 
        & \multirow{2}{*}{Architecture} & \multirow{2}{*}{\# qubit}
        & \multicolumn{3}{c||}{\# Extra \texttt{CNOT} gate}
        & \multicolumn{3}{c||}{Circuit depth} \\ 

        \cline{5-10}
        
        \multirow{2}{*}{} & \multirow{2}{*}{}
        & \multirow{2}{*}{} & \multirow{2}{*}{} 
        & {\frameworkname} & {\qmrname} & {\sabrename}
        & {\frameworkname} & {\qmrname} & {\sabrename}\\
        
        \hline
        
        \multirow{4}{*}{\makecell{$[\![9, 1, 3]\!]$ \\ Surface Code}} & \multirow{4}{*}{2.82} 
        & {Square} & {25} 
        & 0 & 0 & 30
        & 8 & 17 & 16 \\
        \cline{3-10}
        
        \multirow{4}{*}{} & \multirow{4}{*}{} 
        & {Hexagon} & {21}
        & 14 & 27* & 90
        & 19 & 34* & 35 \\
        \cline{3-10}
        
        \multirow{4}{*}{} & \multirow{4}{*}{} 
        & {H-Square} & {29} 
        & 24 & 42* & 90
        & 20 & 49* & 21 \\
        \cline{3-10}

        \multirow{4}{*}{} & \multirow{4}{*}{} 
        & {H-Hexagon} & {46}
        & 56 & 66* & 96
        & 16 & 47* & 26 \\
        
        \hline

        \multirow{4}{*}{\makecell{$[\![16, 4, 3]\!]$ \\ 2D Color Code}} & \multirow{4}{*}{4.29} 
        & {Square} & {35} 
        & 20 & 168* & 270
        & 17 & 132* & 56 \\
        \cline{3-10}
        
        \multirow{4}{*}{} & \multirow{4}{*}{} 
        & {Hexagon} & {50} 
        & 60 & \timeout & 252
        & 18 & \timeout & 47 \\
        \cline{3-10}
        
        \multirow{4}{*}{} & \multirow{4}{*}{} 
        & {H-Square} & {93} 
        & 72 & \timeout & 468
        & 26 & \timeout & 125 \\
        \cline{3-10}

        \multirow{4}{*}{} & \multirow{4}{*}{} 
        & {H-Hexagon} & {115} 
        & 180* & \timeout & 1,038
        & 30* & \timeout & 324 \\
        
        \hline

        \multirow{4}{*}{\makecell{$[\![8, 3, 2]\!]$ \\ 3D Color Code}} & \multirow{4}{*}{3.69} 
        & {Square} & {30}
        & 0 & Not Exist & Not Exist
        & 20 & Not Exist & Not Exist \\
        \cline{3-10}
        
        \multirow{4}{*}{} & \multirow{4}{*}{} 
        & {Hexagon} & {16}
        & 4 & Not Exist & Not Exist
        & 31 & Not Exist & Not Exist \\
        \cline{3-10}
        
        \multirow{4}{*}{} & \multirow{4}{*}{} 
        & {H-Square} & {79}
        & 14 & \timeout & 270
        & 30 & \timeout & 40 \\
        \cline{3-10}

        \multirow{4}{*}{} & \multirow{4}{*}{} 
        & {H-Hexagon} & {55}
        & 38 & \timeout & 228
        & 33 & \timeout & 35 \\

        \hline  
        
        \multirow{4}{*}{\makecell{$[\![7, 1, 3]\!]$ \\ Steane Code}} & \multirow{4}{*}{3.69} 
        & {Square} & {25}
        & 0 & Not Exist & Not Exist
        & 17 & Not Exist & Not Exist \\
        \cline{3-10}
        
        \multirow{4}{*}{} & \multirow{4}{*}{} 
        & {Hexagon} & {21}
        & 0 & Not Exist & Not Exist
        & 18 & Not Exist & Not Exist \\
        \cline{3-10}
        
        \multirow{4}{*}{} & \multirow{4}{*}{} 
        & {H-Square} & {29}
        & 8 & Not Exist & Not Exist 
        & 32 & Not Exist & Not Exist \\
        \cline{3-10}

        \multirow{4}{*}{} & \multirow{4}{*}{} 
        & {H-Hexagon} & {46}
        & 36 & \timeout & 504
        & 30 & \timeout & 112 \\
        
        \hline

        \multirow{4}{*}{\makecell{HGP \\ QLDPC Code}} & \multirow{4}{*}{4.00} 
        & {Square} & {121} 
        & 588* & \timeout & 726
        & 70* & \timeout & 205 \\
        \cline{3-10}
        
        \multirow{4}{*}{} & \multirow{4}{*}{} 
        & {Hexagon} & {128}
        & 876* & \timeout & 1,236
        & 121* & \timeout & 354 \\
        \cline{3-10}
        
        \multirow{4}{*}{} & \multirow{4}{*}{} 
        & {H-Square} & {176}
        & 859* & \timeout & 990
        & 96* & \timeout & 251 \\
        \cline{3-10}

        \multirow{4}{*}{} & \multirow{4}{*}{} 
        & {H-Hexagon} & {265} 
        & 1,275* & \timeout & 1,386
        & 120* & \timeout & 336 \\
        
        \hline
        
        \multirow{4}{*}{\makecell{$[\![81, 1, 9]\!]$ \\ Surface Code}} & \multirow{4}{*}{3.58} 
        & {Square} & {289}
        & 0 & \timeout & 9,594
        & 8 & \timeout & 2,770 \\
        \cline{3-10}
        
        \multirow{4}{*}{} & \multirow{4}{*}{} 
        & {Hexagon} & {225}
        & 6,547* & \timeout & 8,928
        & 391* & \timeout & 2,428 \\
        \cline{3-10}
        
        \multirow{4}{*}{} & \multirow{4}{*}{} 
        & {H-Square} & {251}
        & 288 & \timeout & 9,654
        & 15 & \timeout & 2,458 \\
        \cline{3-10}

        \multirow{4}{*}{} & \multirow{4}{*}{} 
        & {H-Hexagon} & {387}
        & 9,322* & \timeout & 11,064
        & 450* & \timeout & 2,930 \\
        
        \hline
        \end{tabular}
    }
    \label{tab:evaluation}
\end{table*}

\vspace{2pt}
\noindent \textbf{Stabilizer codes:} Our evaluation covers both classical and recent popular stabilizer codes to demonstrate the generality of our framework. For instance, small-scale $[\![9, 1, 3]\!]$ and large-scale $[\![81, 1, 9]\!]$ surface code are selected to show support for the most popular surface code family. The 2D $[\![16, 4, 3]\!]$ and the 3D $[\![8, 3, 2]\!]$ color codes color codes are included to illustrate the influence of distinct code topologies. The classical $[\![7, 1, 3]\!]$ Steane code is selected for its role as a basic block that can be merged into larger stabilizer codes \cite{Meas_no_extra}, such as the $[\![22, 4, 3]\!]$ and $[\![12, 2, 3]\!]$ color codes. The HyperGraph Product (HGP) code \cite{tillich2013quantum} is chosen to demonstrate our framework's applicability to recent popular qLDPC codes.
Moreover, we adopt some manually designed measurement schemes for specific codes to showcase the extensibility of our framework. Specifically, we adapt Shor's scheme \cite{divincenzo2007effective} for the 3D color code and the Steane code to ensure the fault tolerance of the \circuitname~circuits.

To quantify the level of interconnectivity for each stabilizer code, we define the density of a stabilizer code:
\begin{align*}
    Density(C) = \frac{2\sum_{s \in Stabs} wt(s)}{\left| Stabs \right| + \left| Data \right|}
\end{align*}

\vspace{2pt} 
\noindent \textbf{Error Model:} We follow a widely-used circuit-level error model similar to those in \cite{chamberland2020topological, stim, Surf-Stitch}, involving the same error rates and types. Specifically, we apply probabilities ranging from $10^{-3}$ to $10^{-2}$ to the single-qubit depolarizing error channel for single-qubit gates, the two-qubit depolarizing error channel for two-qubit gates, and the Pauli-X error channel for measurement and reset operations. Additionally, for idle errors induced by decoherence, each idle qubit undergoes a single-qubit depolarizing error channel per gate duration with a probability of $2 \times 10^{-4}$ \cite{google2021exponential, Surf-Stitch, das2021adapt}. These errors affect all qubits, including both data and ancilla qubits.

\vspace{2pt} \noindent
\textbf{Metrics:} We consider two key metrics that significantly impact the fidelity of quantum circuit execution and are widely used to evaluate the performance of general QMR method s \cite{OLSQ, Sabre, molavi2022qubit}.
\textit{(1) Extra \texttt{CNOT} gate count:} We evaluate the extra \texttt{CNOT} gate count required by \circuitname~circuits to compensate for sparse architectures. Previous studies \cite{SurfaceCode, stim} and our experiments in \cref{sec:erroranalysis} show the high sensitivity of \circuitname~circuits to two-qubit gate errors, making the \texttt{CNOT} gate count a crucial factor that significantly affects the fidelity of logical qubits.
\textit{(2) Circuit Depth:} We assess the Circuit Depth of entire \circuitname~circuit, which is the maximum time coordinates of all gates. Due to the limitation of current quantum computing technology, physical qubits can only function well up to a short ‘lifetime’. Thus, minimizing depth becomes crucial to ensure efficiency. A larger time-step count would introduce more decoherence errors that accumulate, eventually surpassing the fault tolerance threshold of the error correction procedure. 


\subsection{Compared to Swapping-Based Approaches}
\label{sec:Eval1}
We begin by comparing \frameworkname~with two traditional swapping-based approaches designed for the QMR task of general quantum circuits: \qmrname~(a constraint-based approach) \cite{molavi2022qubit} and \sabrename~(a heuristic approach) \cite{Sabre}. The results in \cref{tab:evaluation} demonstrate that our \frameworkname~consistently synthesizes smaller and faster \circuitname~circuits for stabilizer codes. We highlight the following three results:


\vspace{2pt}
\noindent (1) Our \frameworkname~shows great improvement even when the code synthesis task is small-scale. For instance, in the case of the $[\![9, 1, 3]\!]$ surface code, \frameworkname~demonstrates an average reduction of 26.5\% in the extra \texttt{CNOT} gate count compared to \qmrname~and 74.9\% compared to \sabrename. Additionally, the improvement in circuit depth is even more substantial, with an average reduction of 55.5\% compared to \qmrname~and 34.7\% compared to \sabrename. This highlights the efficiency of \frameworkname's bridging-based approach over the traditional swapping-based approach in addressing connectivity disparity problems in \circuitname~circuit synthesis.

\vspace{2pt}
\noindent (2) The "$*$" symbol and "\timeout" entry indicate that the SAT solver times out, returns a sub-optimal result, or does not produce a result. This often occurs when the code becomes complicated and large, such as the HGP qLDPC code, or when the "code-architecture" gap is substantial, such as mapping the dense $[\![16, 4, 3]\!]$ color code to sparse architectures (Hexagon, Heavy Square, and Heavy Hexagon). This demonstrates that \frameworkname~efficiently prunes the optimization space of \circuitname~circuits by treating and mapping each stabilizer as a whole, rather than mapping each gate in the \circuitname~circuit separately, as traditional QMR approaches do.

\vspace{2pt}
\noindent (3) The "\noexist" entry indicates that the approach verifies the impossibility of finding a workable code synthesis on a device with the given architecture and a limited number of physical qubits, as the number of physical qubits is always restricted on a practical quantum chiplet \cite{gambetta2020ibm, Sycamore, google2023suppressing}.
The inability of swapping-based approaches to leverage the flexibility of overlapping ancilla qubits contributes to the occurrence of "\noexist". These approaches treat data and ancilla qubits as identical, requiring each to be mapped to a distinct physical qubit, which may exceed the available number of physical qubits on the device. For instance, in the case of the $[\![8, 3, 2]\!]$ 3D Color Code and the $[\![7, 1, 3]\!]$ Steane Code, their original \circuitname~circuits apply Shor's scheme and require more than one ancilla qubit for each stabilizer, resulting in a "\noexist" for the swapping-based approaches.


\subsection{Compared to Heuristic Bridging-Based Approach}
\label{sec:Eval2} 
We compare our \frameworkname~with \surfname~\cite{Surf-Stitch}, a heuristic method that can only work on the \textbf{surface code} and \textbf{specific regular architectures}. 
In \cref{tab:evaluation2}, we show the synthesis results of the $d=5$ surface code on both perfect and defective architectures \cite{auger2017fault, lin2024codesign, suzuki2022q3de} that have a faulty qubit in the middle due to fabrication defects. The result shows that \frameworkname~has a wider range of applicability while achieving equal or better \circuitname~circuits than \surfname. We highlight the following three results:

\begin{table}[!h]
    \centering
    \caption{Detailed information about synthesized distance-5 surface codes on perfect or defective architectures.}
    \resizebox{0.45\textwidth}{!}{
        \begin{tabular}{|c|c|c|c|c|c|} 
        
        \hline
        \multicolumn{2}{|c|}{\multirow{3}{*}{\diagbox[innerwidth=\widthof{AAAAAAAAAAAAA}]{Architecture}{Metric}}}  
        & \multicolumn{2}{c|}{\frameworkname} 
        & \multicolumn{2}{c|}{\surfname} \\ 
        
        \cline{3-6}
        \multicolumn{2}{|c|}{\multirow{3}{*}{}} 
        & \multirow{2}{*}{\makecell{\# Extra \\ \texttt{CNOT} gate}}
        & \multirow{2}{*}{\makecell{~Circuit~ \\ depth}} 
        & \multirow{2}{*}{\makecell{\# Extra \\ \texttt{CNOT} gate}} 
        & \multirow{2}{*}{\makecell{~Circuit~ \\ depth}} \\ 

        \multicolumn{2}{|c|}{\multirow{3}{*}{}} 
        & \multirow{2}{*}{} 
        & \multirow{2}{*}{} 
        & \multirow{2}{*}{}
        & \multirow{2}{*}{}\\
        
        \hline
        
        \multirow{2}{*}{Square} & {Perfect} 
        & 0 & 8 & 0 & 8\\
        \multirow{2}{*}{} & {Defective} 
        & 12 & 17 & $\times$ & $\times$\\
        
        \hline

        \multirow{2}{*}{Hexagon} & {Perfect} 
        & 72 & 20 & 120 & 26\\
        \multirow{2}{*}{} & {Defective} 
        & 76 & 17 & $\times$ & $\times$\\
        
        \hline

        \multirow{2}{*}{\makecell{Heavy \\ Square}} & {Perfect}
        & 80 & 15 & 80 & 24\\
        \multirow{2}{*}{} & {Defective} 
        & 94 & 20 & $\times$ & $\times$\\
        
        \hline

        \multirow{2}{*}{\makecell{Heavy \\ Hexagon}} & {Perfect}
        & 216 & 18 & 224 & 40 \\
        \multirow{2}{*}{} & {Defective} 
        & 216 & 23 & $\times$ & $\times$\\
        
        \hline
        \end{tabular}
    }
    \label{tab:evaluation2}
\end{table}

\vspace{3pt}
\noindent (1) Although at most cases of perfect architectures, the improvement on \texttt{CNOT} gates number by \frameworkname~ is marginal, as the \surfname's mannual mapping of surface code is already near optimal, \frameworkname~still find a synthesis on hexagon with using 40\% less extra \texttt{CNOT} gates. Moreover, due to our more concurrent scheduling strategy, \frameworkname~achieves a average reduction of 38.5\% in circuit depth. 

\vspace{3pt}
\noindent (2) In the presence of faulty physical qubits that cannot be used in \circuitname~circuits due to fabrication defects \cite{auger2017fault, nagayama2017surface, lin2024codesign}, \surfname~fails to operate because the architecture is no longer regular. However, our \frameworkname~adapts seamlessly to these defective architectures with only a minor increase in \texttt{CNOT} gates and circuit depth.

\begin{figure}[!ht]
    \centering
    \includegraphics[width=0.45\textwidth]{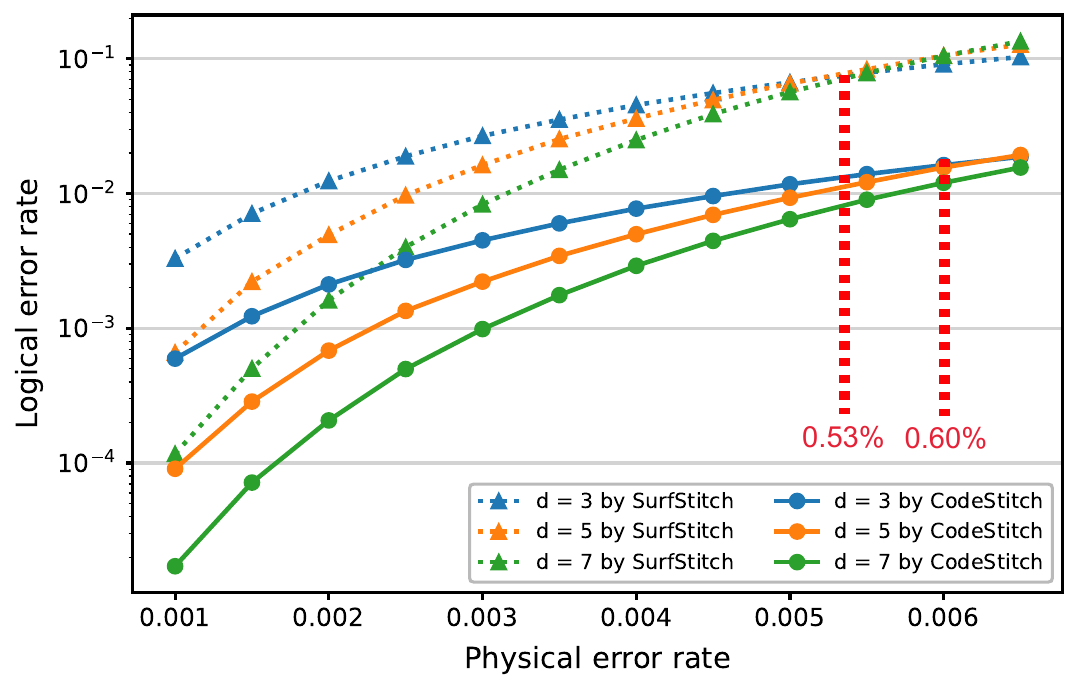}
    \caption{Logical error rate of distance-5 surface codes synthesized by \frameworkname~and \surfname~on heavy-square architecture. The error threshold is the physical error rate at which the code curves of different distances intersect, which is represented with red dotted lines.}
    \label{fig:threshold}
\end{figure}

\vspace{3pt}
\noindent (3) For two synthesized \circuitname~circuits with an equal count of \texttt{CNOT} gates (e.g. perfect heavy-square), circuit depth significantly affects their error correction capability.
As shown in \cref{fig:threshold}, we simulated the logical error rates for surface codes of different distances on the heavy-square architecture, each using $10^6$ slots of simulation. The results for \frameworkname~consistently show significantly lower logical error rates, ranging from 6.8x to 8.6x lower for the distance-7 surface code, despite both approaches using the same number of \texttt{CNOT} gates.
Moreover, surface codes synthesized by \frameworkname~also show an improved error threshold. Although the improvement from 0.53\% to 0.60\% may seem marginal, it significantly affects the scalability of surface codes. For instance, if we aim to implement a logical qubit on a device with a physical error rate of 0.50\%, \surfname~would require 10x more physical qubits to achieve the same logical error rate.

\subsection{Ablation Study}
To justify prioritizing "\# extra \text{CNOT} gates" over "circuit depth" as the primary optimization goal and employing a two-stage SAT approach to optimize them separately, we perform a \textit{breakdown analysis} of the logical error rate to identify the most critical factor for the fidelity of the \circuitname~circuit. Additionally, we conduct a \textit{scalability analysis} to demonstrate the performance of our heuristic relaxation method.

\vspace{3pt}
\noindent \textbf{Breakdown analysis of logical error rate.}
\label{sec:erroranalysis}
This study provides a breakdown analysis of how the error threshold shifts across two vital error types: two-qubit \texttt{CNOT} gate errors and device idle errors. 
The simulation focuses on surface codes on 2-D grid hardware, but the conclusion can be readily extrapolated to broader contexts.
This analysis provides additional support for the rationale behind prioritizing the reduction of \# \texttt{CNOT} gates (linked to accumulated \texttt{CNOT} errors) over circuit depth (linked to accumulated idle errors) when optimizing QEC code synthesis. 

\begin{figure}[h!]
    \centering
    \includegraphics[width=0.48\textwidth]{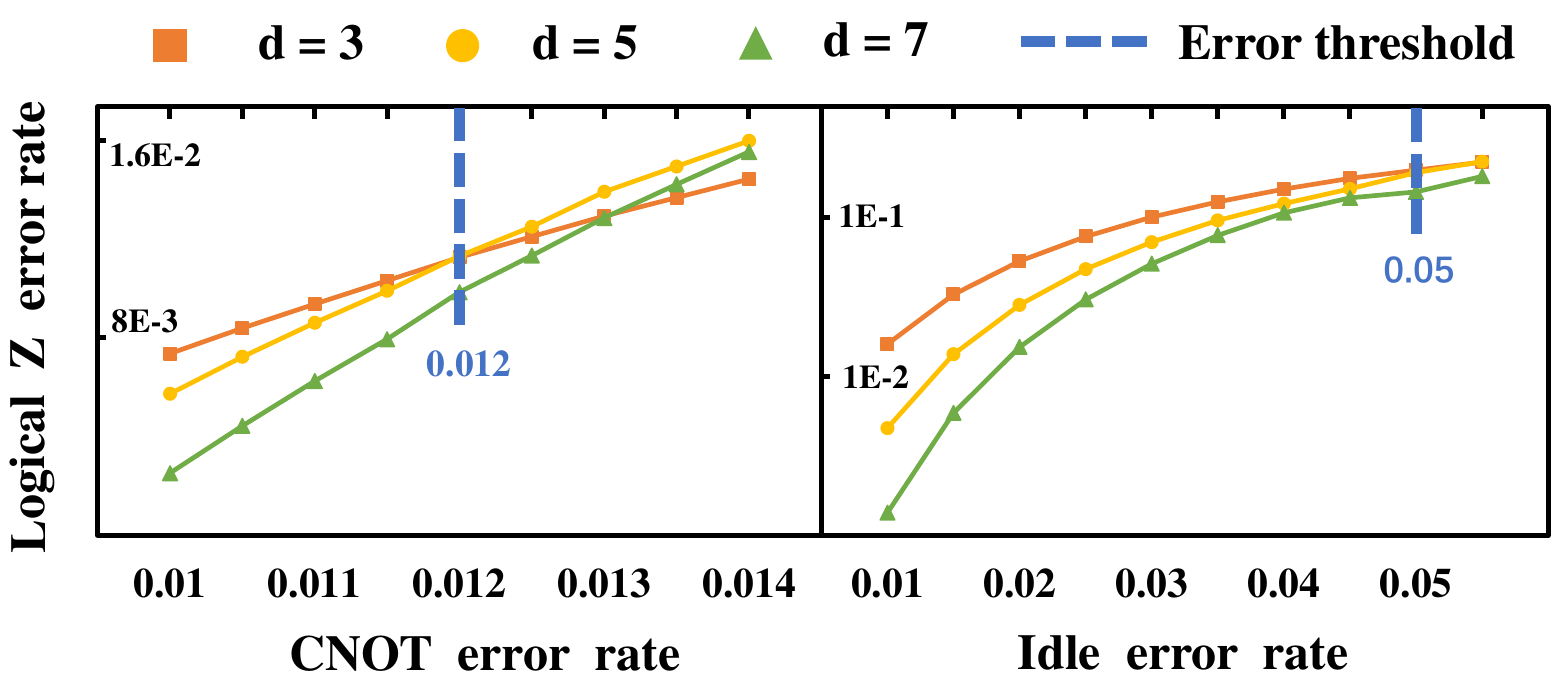}
    \caption{Breakdown analysis of the influence of \texttt{CNOT} gate errors and idle errors on the logical error rate.}
    \label{fig:StabCode}
\end{figure}

We conducted $10^6$ simulation slots for each logical error point. When analyzing the \texttt{CNOT} gate error, the idle error is set to 0, and vice versa. As shown in \cref{fig:StabCode}, pure \texttt{CNOT} gate error exhibits a per-step error threshold of 1.2\%, while pure idle error demonstrates a threshold of 5\%. The significantly lower threshold for \texttt{CNOT} errors highlights both the reduced tolerance for this type of error and its amplified impact on the overall logical error rate. This emphasizes the importance of prioritizing the reduction of the \texttt{CNOT} gate count.

\vspace{3pt}
\noindent\textbf{Scalability Analysis.}
\label{sec:scalability analysis}
In this study, we evaluate the scalability of \frameworkname~in comparison to the solver-based \qmrname~and test our relaxation method by code partition on dense and large-scale stabilizer codes. 

\begin{table}[!h]
    \centering
    \caption{Size of MaxSAT-encoded problem for $[\![9, 1, 3]\!]$ surface code synthesis by \frameworkname~and \qmrname}
    \resizebox{0.45\textwidth}{!}{
        \renewcommand{\arraystretch}{1.5}
        
        \begin{tabular}{|c|c|*{4}{c|}} 
        \hline
        \multicolumn{2}{|c|}{\multirow{2}{*}{\diagbox[innerwidth=\widthof{AAAAAAAAAAAAAA}]{Metrics}{Arch}}} 
        & \multirow{2}{*}{\makecell{Square}}
        & \multirow{2}{*}{\makecell{Hexagon}} 
        & \multirow{2}{*}{\makecell{Heavy \\ Square}} 
        & \multirow{2}{*}{\makecell{Heavy \\ Hexagon}} \\ 
        
        \multicolumn{2}{|c|}{\multirow{2}{*}{}} 
        & \multirow{2}{*}{} 
        & \multirow{2}{*}{} 
        & \multirow{2}{*}{}
        & \multirow{2}{*}{}\\
        
        \hline
        
        \multirow{2}{*}{\# Variable} & {\frameworkname} 
        & $6.38 \times 10^3$ & $1.24 \times 10^4$
        & $2.27 \times 10^4$ & $1.07 \times 10^5$ \\
        \multirow{2}{*}{} & {\qmrname} 
        & $5.52 \times 10^4$ & $4.04 \times 10^4$ 
        & $7.24 \times 10^4$ & $1.71 \times 10^5$ \\
        
        \hline

        \multirow{2}{*}{\makecell{\# Hard \quad \\ clause}} & {\frameworkname} 
        & $1.51 \times 10^4$ & $2.64 \times 10^4$
        & $4.76 \times 10^4$ & $2.18 \times 10^5$ \\
        \multirow{2}{*}{} & {\qmrname} 
        & $1.80 \times 10^6$ & $9.98 \times 10^5$
        & $1.83 \times 10^6$ & $4.22 \times 10^6$ \\
        
        \hline
        
        \multirow{2}{*}{\makecell{\# Soft \quad \\ clause}}& {\frameworkname} 
        & $228$ & $196$  
        & $260$ & $396$ \\
        \multirow{2}{*}{} & {\qmrname} 
        & $14,400$ & $10,080$  
        & $19,488$ & $49,680$ \\
        \hline
        \end{tabular}
    }
    \vspace{-3pt}
    \label{tab:scalabilty}
\end{table}

The sizes of the MaxSAT-encoded problem are shown in \cref{tab:scalabilty}.
It reveals that tasks from \qmrname~exhibit 1.5x to 8.5x more variables, 20x to 120x more hard constraints, and 50x to 125x more soft constraints than the tasks from \frameworkname. This indicates that our \frameworkname~method offers a more efficient approach to encoding the QEC code mapping problem than \qmrname, by formalizing the special structure characteristic of the measurement circuit in stabilizers. 

Moreover, the synthesis results of the HGP qLDPC code and the $d=9$ surface code in \cref{tab:evaluation} further demonstrate the efficiency of \frameworkname~on dense and large-scale stabilizer codes using the heuristic relaxation method. It shows significant improvements in reducing extra \texttt{CNOT} gates and circuit depth compared to traditional swapping-based approaches.

%% file: 07_related_work.tex
\section{Related Work}

\vspace{3pt}
\noindent\textbf{Swapping-based methods. }Extensive research has addressed the challenge of mapping general quantum circuits onto hardware with limited connectivity. These methods typically fall into one of three categories: constraint-based~\cite{wille2019mapping, murali2019noise, tan2020optimal, molavi2022qubit}, heuristic~\cite{Sabre, zulehner2018efficient, zhang2021time}, or a hybrid approach~\cite{finigan2018qubit}. However, these techniques are inadequate for \circuitname~circuits due to the inserted SWAP gates, which disrupt qubit locations and hinder parallel execution of stabilizer measurements with shared data qubits. Moreover, introducing SWAP gates can elevate the circuit’s error rate, compromising the precision needed for effective \circuitname~circuits.

\vspace{3pt}
\noindent\textbf{Bridging-based methods. }The original bridging-based methods \cite{Meas_no_extra, Flag-Bridge, chao2020flag, chao2018fault, chamberland2018flag} extend single ancilla qubits into ancilla bridges to alleviate the connectivity constraint. This approach offers fixed qubit locations and introduces fewer additional gates compared to swap-based methods, enhancing performance. However, these methods are designed for single \circuitname~circuits and struggle with complex dependencies among stabilizer measurements within QEC codes. Recent efforts have applied bridging-based techniques to entire QEC codes \cite{tremblay2022constant, chamberland2020topological, Surf-Stitch}, but they are limited to specific QEC codes and hardware architectures, lacking adaptability to broader settings. Furthermore, these approaches rely on heuristic design without optimality guarantees.

%% file: 08_conclusion.tex
\section{Conclusion}

This paper introduces the first automated compilation framework for implementing \circuitname~circuits using the bridging method. Our approach demonstrates broad applicability across diverse QEC codes and hardware architectures, including systems with defective qubits. By systematically classifying and leveraging the primary flexibilities of the bridging method, we effectively explore its extensive design space and formalize \circuitname~circuit implementation as a two-stage MaxSAT problem solved by high-performance SAT solvers. Comparative evaluations show our method significantly outperforms existing swap-based and bridging approaches while achieving superior adaptability. These advancements represent a significant step toward realizing long-term fault-tolerant quantum computing goals.

%% file: ACK.tex
\section{Acknowledgments}
We thank the anonymous reviewers for their valuable and constructive feedback. This work was supported in part by NSF 2138437, NSF 2048144, and the Robert N. Noyce Trust. 
This material is based upon work supported by the U.S. Department of Energy, Office of Science, National Quantum Information Science Research Centers, Quantum Science Center.